\documentclass[12pt,letterpaper]{article}

\usepackage{times}
\usepackage{amsmath,amsfonts}
\usepackage{graphicx}
\usepackage{setspace}
\usepackage{indentfirst}
\usepackage{url}

\usepackage{color}
\definecolor{hilit}{RGB}{17,60,140}

\usepackage{hyperref}
\hypersetup{pdfpagemode=UseNone} 
\hypersetup{hidelinks}

\usepackage[authoryear]{natbib}
\bibpunct{(}{)}{;}{a}{}{,}

\newenvironment{hanging}
{\begin{list}{}
        {\setlength{\labelwidth}{0in}
         \setlength{\leftmargin}{1em}
         \setlength{\itemindent}{-1em}
        }
}
{\end{list}}

\begin{document}
\setstretch{2.0}

\vspace*{8mm}
\begin{center}

\textbf{\Large A simple regression-based method to map \\
  quantitative trait loci underlying function-valued
  phenotypes}

\bigskip \bigskip \bigskip \bigskip

{\large Il-Youp Kwak$^*$, Candace R. Moore$^\dagger$, Edgar
  P. Spalding$^\dagger$, Karl W. Broman$^{\ddagger,1}$}

\bigskip \bigskip

Departments of $^*$Statistics, $^\dagger$Botany, and $^\ddagger$Biostatistics and Medical
Informatics, \\
University of Wisconsin--Madison, Madison, Wisconsin 53706
\end{center}

\vfill

\hfill
{\footnotesize 15 May 2014}

\newpage

\noindent \textbf{Running head:} Mapping QTL for function-valued traits

\bigskip \bigskip \bigskip


\bigskip \bigskip \bigskip

\noindent \textbf{$^1$Corresponding author:}

\begin{tabular}{lll}
 \\
 \hspace{1cm} & \multicolumn{2}{l}{Karl W Broman} \\
 & \multicolumn{2}{l}{Department of Biostatistics and Medical Informatics} \\
 & \multicolumn{2}{l}{University of Wisconsin--Madison} \\
 & \multicolumn{2}{l}{2126 Genetics-Biotechnology Center} \\
 & \multicolumn{2}{l}{425 Henry Mall} \\
 & \multicolumn{2}{l}{Madison, WI 53706} \\
 \\
 & Phone: & 608--262--4633 \\
 & Email: & \verb|kbroman@biostat.wisc.edu|
\end{tabular}

\newpage

\centerline{\sffamily \textbf{Abstract}}
Most statistical methods for QTL mapping focus on a single phenotype.
However, multiple phenotypes are commonly measured, and recent
technological advances have greatly simplified the automated
acquisition of numerous phenotypes, including function-valued
phenotypes, such as growth measured over time. While there exist
methods for QTL mapping with function-valued phenotypes, they are
generally computationally intensive and focus on single-QTL models. We
propose two simple, fast methods that maintain high power and precision
and are amenable to extensions with multiple-QTL models using
a penalized likelihood approach.  After
identifying multiple QTL by these approaches, we can view the
function-valued QTL effects to provide a deeper understanding of the
underlying processes.  Our methods have been implemented as a package
for R, funqtl.

\newpage

\centerline{\sffamily \textbf{Introduction}}

There is a long history of work to map genetic loci (called
quantitative trait loci, QTL) influencing
quantitative traits. Most statistical methods for QTL mapping,
such as interval mapping \citep{Lander1989},
focus on a single phenotype.
However, multiple phenotypes are commonly measured, and recent
technological advances have greatly simplified the automated
acquisition of numerous phenotypes, including phenotypes measured over
time. Phenotypes measured over time, an example of a
function-valued trait, have a number of advantages, including the
ability to dissect the time course of QTL effects.

A simple and intuitive approach to the analysis of such data is to
perform QTL analysis
at each time point, individually, to identify QTL that affect the
phenotype at each time point. This method is simple, however it
does not consider the smooth association across time points, and so it may
have less power to detect QTL. Moreover, it can be difficult to
combine the results across time
points into a consistent story.

A second approach is to fit parametric curves to the data from each
individual and treat the parameter estimates as phenotypes in QTL
analysis \citep[e.g., see][]{Kendziorski2002}. \citet{Ma2002} expanded this approach by fitting
a logistic growth model, $g(t) = \frac{a}{1+b e^{-rt}}$, at each
putative QTL position, with parameters depending on QTL genotype.
This approach can have
high power if the model is correct, but it can be difficult to
interpret the results if QTL have pleiotropic effects on multiple
parameters, and the parameters may have no obvious biologic or
mechanistic interpretation.

Another natural approach is to use a non-parametric method so that we don't need
to specify the functional shape. For example, \citet{Yang2009}
proposed a non-parametric functional QTL mapping method that used a certain number
of basis functions to fit a function-valued phenotype. For example,
we might use ten basis functions. This reduces the dimension from the number of
time points to ten, and this is done in a flexible way, guided by the
data. \citet{Min2011} extended this method to multiple-QTL models
using Markov chain Monte Carlo (MCMC) techniques.

\citet{Xiong2011} proposed an additional non-parametric functional mapping
method based on estimating equations (EE). This method is fast
and allows the selection of multiple QTL by a test statistic that they
proposed. \citet{Sillanpaa2012} proposed another Bayesian multiple-QTL
mapping method based on hierarchical modeling.

Important limitations of existing approaches for the
analysis of function-valued traits are that they focus on single-QTL
models or exhibit slow speed in multiple-QTL search. We describe two simple
methods for QTL mapping with function-valued traits and, following the
approach of \citet{Broman2002} and \citet{Manichaikul2009}, extend them
for the consideration of multiple-QTL models.

We investigate the performance of our approach in computer simulations
and apply it to data on a plant growth response known as root
gravitropism, which \citet{Moore2013} measured by automated image analysis over a
time course of 8 hours across a population of \emph{Arabidopsis
thaliana\/} recombinant inbred lines (RIL).
Our aim is to identify the
genetic loci (QTL) that influence the function-valued phenotype, and
to characterize their effects over time.

\clearpage

\centerline{\sffamily \textbf{Methods}}

We will focus on the case of recombinant inbred lines (RIL). Two
inbred strains, say A and B, are crossed and then the F$_1$
hybrids are subjected to either selfing or sibling mating for many
generations to create a new inbred line whose genome is a mosaic of
the A and B genomes. This is done multiple times in parallel. At any
genomic position, the RIL are homozygous AA or BB.

\textbf{Single-QTL analysis:}
The most popular method for QTL mapping is interval mapping, developed by
\citet{Lander1989}. Consider a single phenotype,
$y$, and assume there is one QTL, with the model
$y = \mu + \beta q + \epsilon$
where $q$ denotes the QTL genotype, taking the value 0 for genotype AA and 1
for genotype BB, and $\epsilon \sim \text{N}(0,\sigma^2)$. Thus $\mu$ is the
average phenotype for QTL genotype AA and
$\beta$ is the effect of the QTL.

A key problem is that genotypes are observed only at markers, and we
wish to consider positions between markers as putative QTL
locations. However, we may calculate $p = \Pr(q=\text{BB} | \text{marker data})$.
The phenotype, given the marker data, then follows a mixture
of normal distributions with known mixing proportion, $p$. An EM algorithm
(Dempster et al, 1977) may be used to derive maximum likelihood
estimates of the three parameters, $\mu, \beta$ and $\sigma$. This is
done at each putative QTL location, $\lambda$. Alternatively, one may
use regression of $y$ on $p$ to provide a fast approximation \citep{Haley1992}.

\citet{Lander1989} summarized the evidence for a QTL at
position $\lambda$ by the LOD score, LOD$(\lambda)$, which is the log$_{10}$
likelihood ratio comparing the hypothesis of a single QTL at
position $\lambda$ to the null hypothesis of no QTL.
LOD scores indicate evidence of presence of QTL. To assess the
statistical significance of the results, one must deal with the
multiple hypothesis testing issue, from the scan across the
genome. This is best handled by a permutation test \citep{Churchill1994}.

With a function-valued trait, $y(t)$, the model becomes
$y(t) = \mu(t) + \beta(t) q + \epsilon(t)$.
(We focus on the case of a phenotype measured over time, but the
approach may be applied to any function-valued trait of a single
parameter, such as a dose-response curve, or really to any
multivariate trait.)
The simplest approach is to apply single-QTL analysis for each time
$t$, individually. This gives LOD($t, \lambda$) for time $t$ at QTL position
$\lambda$. We seek to integrate the information across time points to give
overall evidence for QTL. Two simple rules are to take the average or maximum LOD
scores across times, respectively:
\begin{equation*}
  \text{SLOD}(\lambda) = \frac{1}{T} \sum_{t=1}^{T} \text{LOD}(t,\lambda)
\end{equation*}
\begin{equation*}
  \text{MLOD}(\lambda) = \max_t \text{LOD}(t,\lambda),
\end{equation*}
where $T$ is the number of time points.

With MLOD, one asks whether there is any time point at which a locus
has an effect, while SLOD concerns the overall effect of the
locus. MLOD will be more powerful for identifying QTL with large
effects over a brief interval of time, while SLOD will be more
powerful for identifying loci with effect over a large interval.

To assess significance, we permute the rows in the phenotype matrix
relative to rows in the genotype
matrix, calculate the statistic across genome, and record the
maximum. We take the 95th percentile of the genome-wide maxima as a
5\% significance threshold.

Rapid computations are enabled by the simultaneous analysis of the multiple time
points. Whereas coefficient estimates at a single time point would be
obtained as $\hat{\beta} = (X'X)^{-1} X' y$, with multiple time points
we may replace the vector $y$ with a matrix $Y$, whose columns
correspond to the multiple time points. This gives
$\hat{\boldsymbol{\beta}} = (X'X)^{-1} X' Y$. The matrix inversion is
performed once at each putative QTL position, and the simultaneous
analysis of multiple time points is obtained by matrix
multiplication, and so the computations are linear in the number of
time points.

\textbf{Multiple-QTL analysis:}
\citet{Broman2002} developed a method to find multiple QTL in
an additive model by using a penalized LOD score criterion,
$\text{pLOD}_a(\gamma) = \text{LOD}(\gamma) - \text{T}|\gamma|$,
where $|\gamma|$ is the number of QTL in a model $\gamma$, and $T$ is a
penalty constant, chosen as the $1 - \alpha$ quantile of the
genomewide maximum LOD score under the null hypothesis of no QTL,
derived from a permutation test.

The approach is readily extended to the function-valued case, by
replacing the LOD score for a model with SLOD or
MLOD, to integrate the information across time points. The penalty,
$T$, is the $1-\alpha$ significance threshold from a
single-QTL genome scan, derived using the permutation procedure
described above.

To search the space of models, we use the stepwise model search
algorithm of \citet{Broman2002}: we use forward selection up to a
model of fixed size (e.g., 10 QTL), followed by backward elimination
to the null model. The selected model $\hat{\gamma}$ is that which maximizes the
penalized SLOD or MLOD criterion, among all models visited.

The selected model is of the form
$y(t) = \hat{\mu}(t) + \sum_j \hat{\beta}_j(t) q_j + \epsilon(t)$,
where the $q_j$ are selected QTL (taking value 0 for genotype AA and 1
for genotype BB), $\hat{\mu}(t)$
is an estimated baseline
function, and $\hat{\beta}_j(t)$ is the estimated effect of QTL $j$ at
time $t$.

\clearpage

\centerline{\sffamily \textbf{Application}}

As an illustration of our approaches, we considered data from
\citet{Moore2013} on gravitropism in Arabidopsis recombinant inbred
lines (RIL), Cape Verde Islands (Cvi) $\times$ Landsberg erecta (Ler).
For each of 162 RIL, 8--20 replicate seeds per line were germinated and then
rotated 90 degrees, to change the orientation of gravity. The growth
of the seedlings was captured on video, over the course of eight
hours, and a number of phenotypes were derived by automated image
analysis.

We focus on the angle of the root tip, in degrees, over time (averaged
across replicates within an RIL), and consider only the first of
two replicate data sets examined in \citet{Moore2013}. There is
genotype data at 234 markers on five chromosomes; the function-valued
root tip angle trait was measured at 241 time points (every two minutes
for eight hours). The estimated genetic map and the trait values for
five randomly selected RIL are displayed in Figure~S1. The average and
SD of the root tip angle at the individual time points, and the correlations between time
points, are displayed in Figure~S2.

The data are available at the QTL Archive,
\href{http://qtlarchive.org/db/q?pg=projdetails&proj=moore_2013b}{
\tt http://qtlarchive.org/db/q?pg=projdetails\&proj=moore\_2013b}.

\begin{figure}[!ht]
\begin{center}
\includegraphics{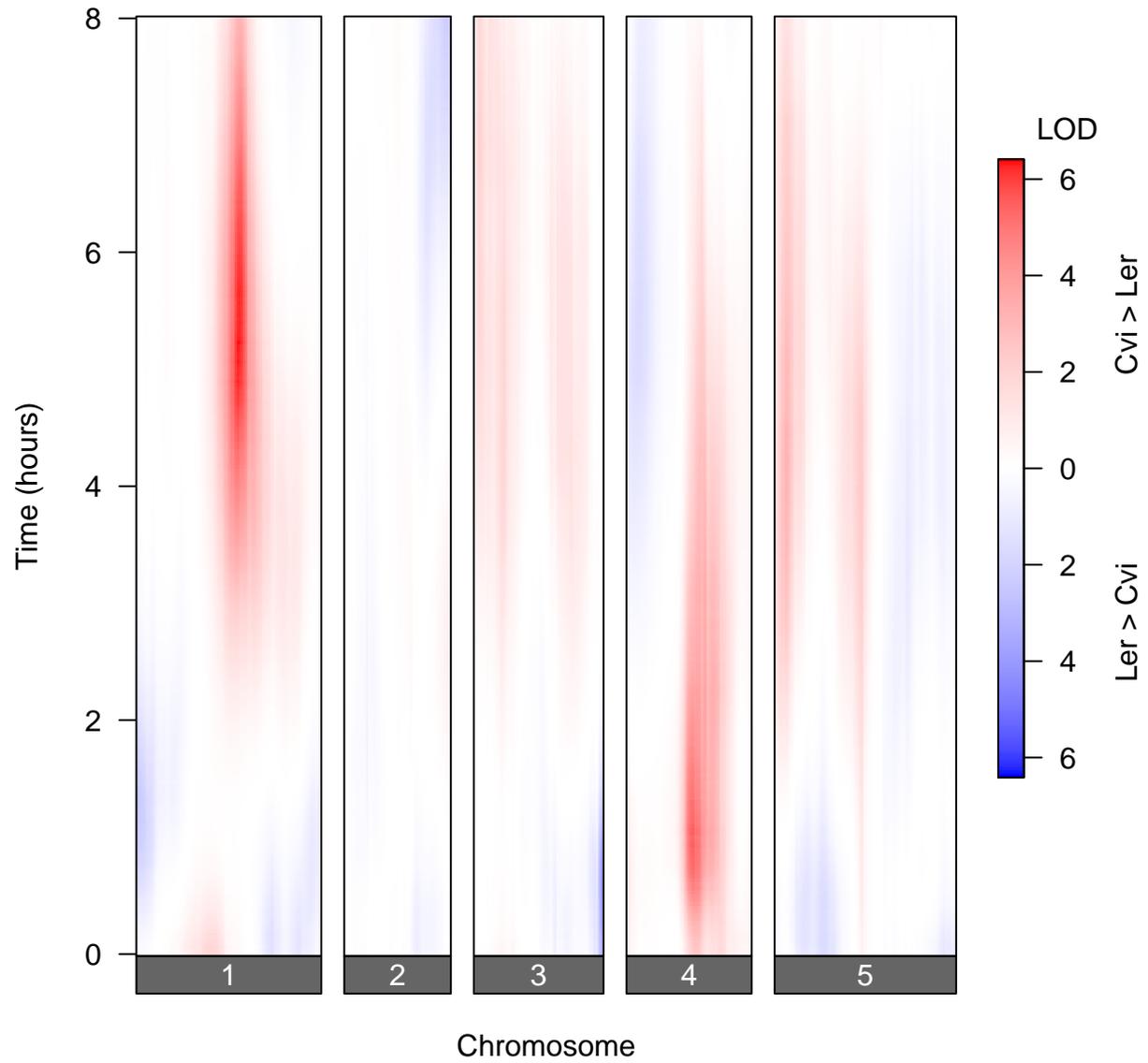}
\vspace{1cm}
 \caption{\label{lodimage1} Signed LOD scores from single-QTL genome
   scans, with each time point considered individually.}
\end{center}
\end{figure}

\textbf{Single-QTL analysis:}
We first applied interval mapping by Haley-Knott regression
\citep{Haley1992}, considering each time point individually.
The results are displayed in Figure~\ref{lodimage1}, with the x-axis
representing genomic position and the y-axis representing time, and so
each horizontal slice is a genome scan for one time point. We plot a
signed LOD score, with the sign representing the estimated direction
of the QTL effect: red indicates that lines with the Cvi allele had higher
phenotype than the lines with the Ler allele; blue indicates that
lines with the Ler allele had higher phenotype than the lines with the
Cvi allele.

The most prominent QTL are on chromosomes 1 and 4; in both cases the
Cvi allele had higher phenotype than the Ler allele.  The chromosome
1 QTL affects later times, and the chromosome 4 QTL affects earlier
times. There is an additional QTL of interest on distal chromosome 3,
with the Ler allele having higher phenotype at early times.

\begin{figure}[!ht]
\begin{center}
\includegraphics{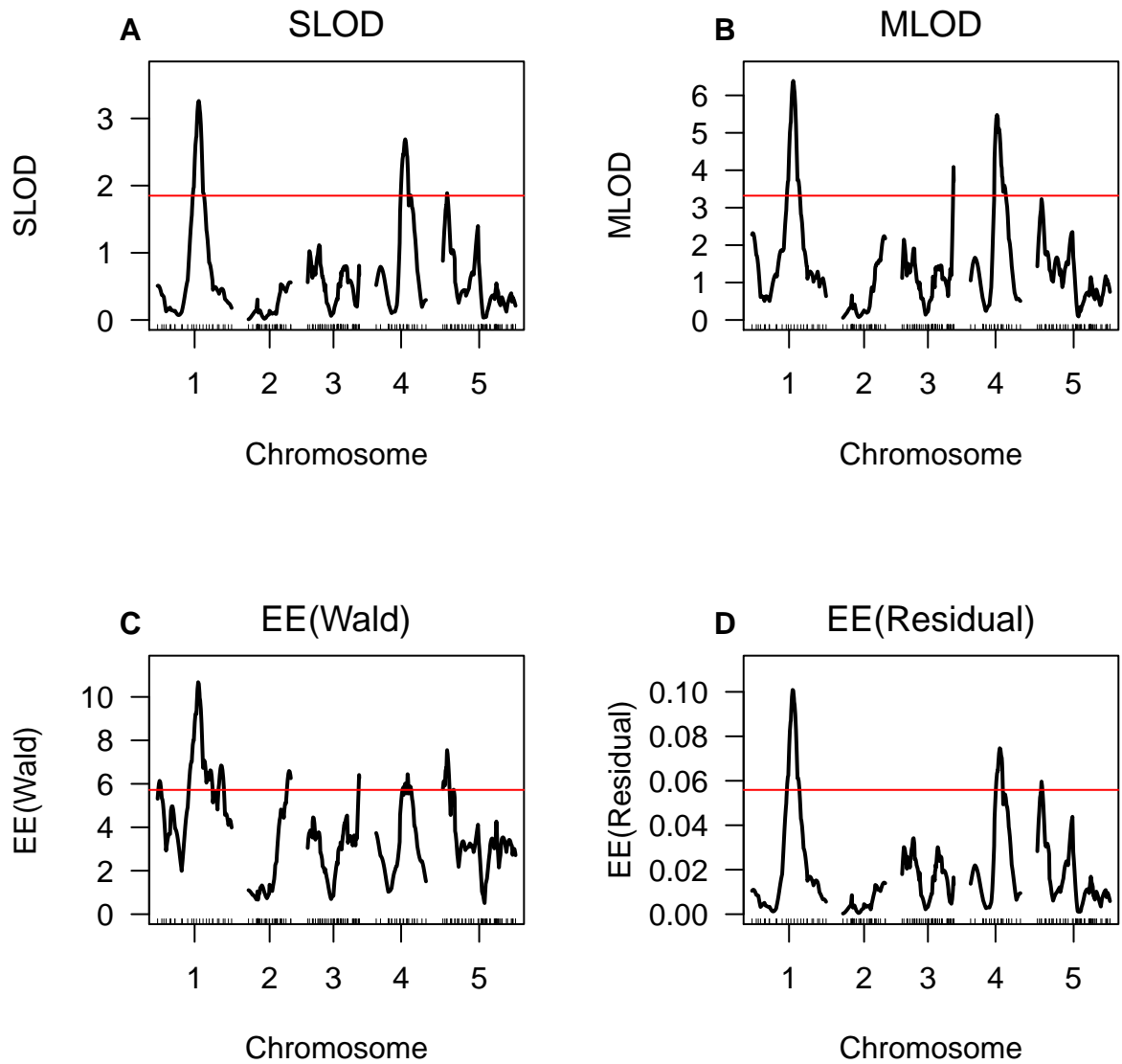}
\vspace{1cm}
 \caption{\label{plod1} The SLOD, MLOD, EE(Wald) and EE(Residual)
     curves for the root tip angle data. A red
     horizontal line indicates the calculated 5\% permutation-based threshold.}
\end{center}
\end{figure}

The SLOD and MLOD statistics combine the results across time points,
by taking the average or the maximum LOD, respectively, at each genomic
location. The results are in Figure~\ref{plod1}A and \ref{plod1}B. Horizontal lines
indicate the 5\% genome-wide significance thresholds, derived by a permutation test.

We also applied the estimating equations approach of
\citet{Xiong2011}. This has two variants: a Wald statistic, denoted
EE(Wald), and a residual error statistic, denoted
EE(Residual). Results are displayed in Figure~\ref{plod1}C and
\ref{plod1}D, again with horizontal lines indicating the 5\%
genome-wide significance thresholds.

The 5\% significance thresholds for the four methods, derived from
permutation tests with 1000 permutation replicates, are shown in
Table~S1.

All four methods identify QTL on chromosomes 1, 4 and 5. The MLOD and
EE(Wald) methods further identify a QTL on chromosome 3, and
the EE(Wald) method identifies a further QTL on chromosome 2.

\textbf{Multiple-QTL analysis:}
Methods that account for multiple QTL may improve power and better separate evidence for
linked QTL. We extended the approach of \citet{Broman2002} for
function-valued traits. Here we focus on additive QTL models, and
extend the SLOD and MLOD statistics.

The penalized-SLOD criterion, with the 5\% significance threshold as
the penalty, indicated a two-QTL model with QTL on chromosomes 1
(at 60~cM) and 4 (at 43~cM).
The penalized-MLOD statistic indicated a
three-QTL model, with an additional QTL on chromosome 3
(at 76.1~cM).
The positions of the QTL on chromosomes 1 and 4 were changed slightly
relative to the inferred QTL model by the penalized-SLOD criterion;
with the penalized-MLOD criterion, the chromosome 1 QTL was at 62~cM
and the chromosome 4 QTL was at 39~cM.

\begin{figure}[!ht]
\begin{center}
\includegraphics{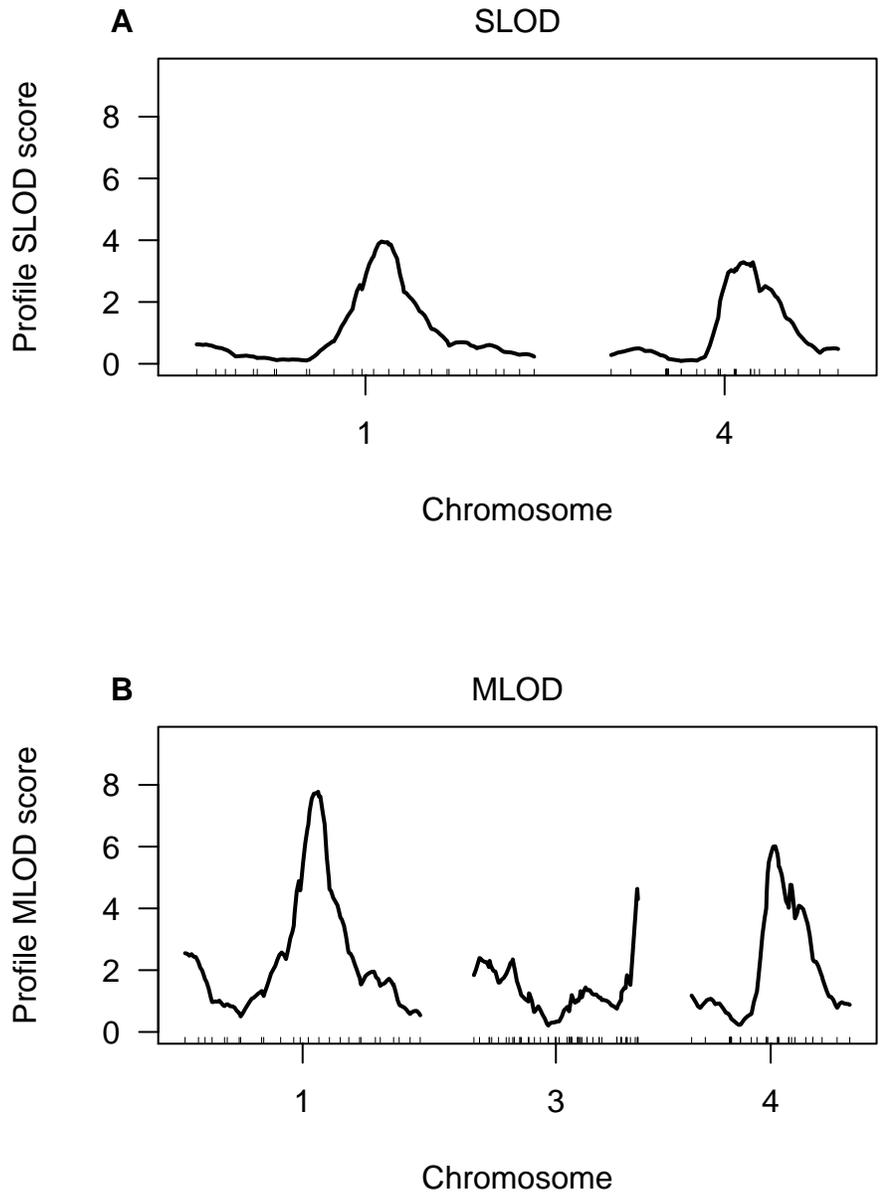}
\vspace{1cm}
 \caption{\label{pro1} SLOD and MLOD profiles for a multiple-QTL model
   with the root tip angle data set.}
\end{center}
\end{figure}

Following an approach developed by \citet{Zeng2000}, we derived
profile log likelihood curves, to visualize the evidence and localization
of each QTL in the context of a multiple-QTL model: The position of each QTL was varied
one at a time, and at each location for a given QTL, we
derived a LOD score comparing the multiple-QTL model
with the QTL under consideration at a particular position and the
locations of all other QTL fixed,
to the model with the given QTL omitted.
This profile is calculated for each time point, individually, and then
the SLOD (or MLOD) profiles are obtained by averaging (or maximizing)
across time points. The SLOD and MLOD profiles are shown in Figure~\ref{pro1}.

\begin{figure}[!ht]
\begin{center}
\includegraphics{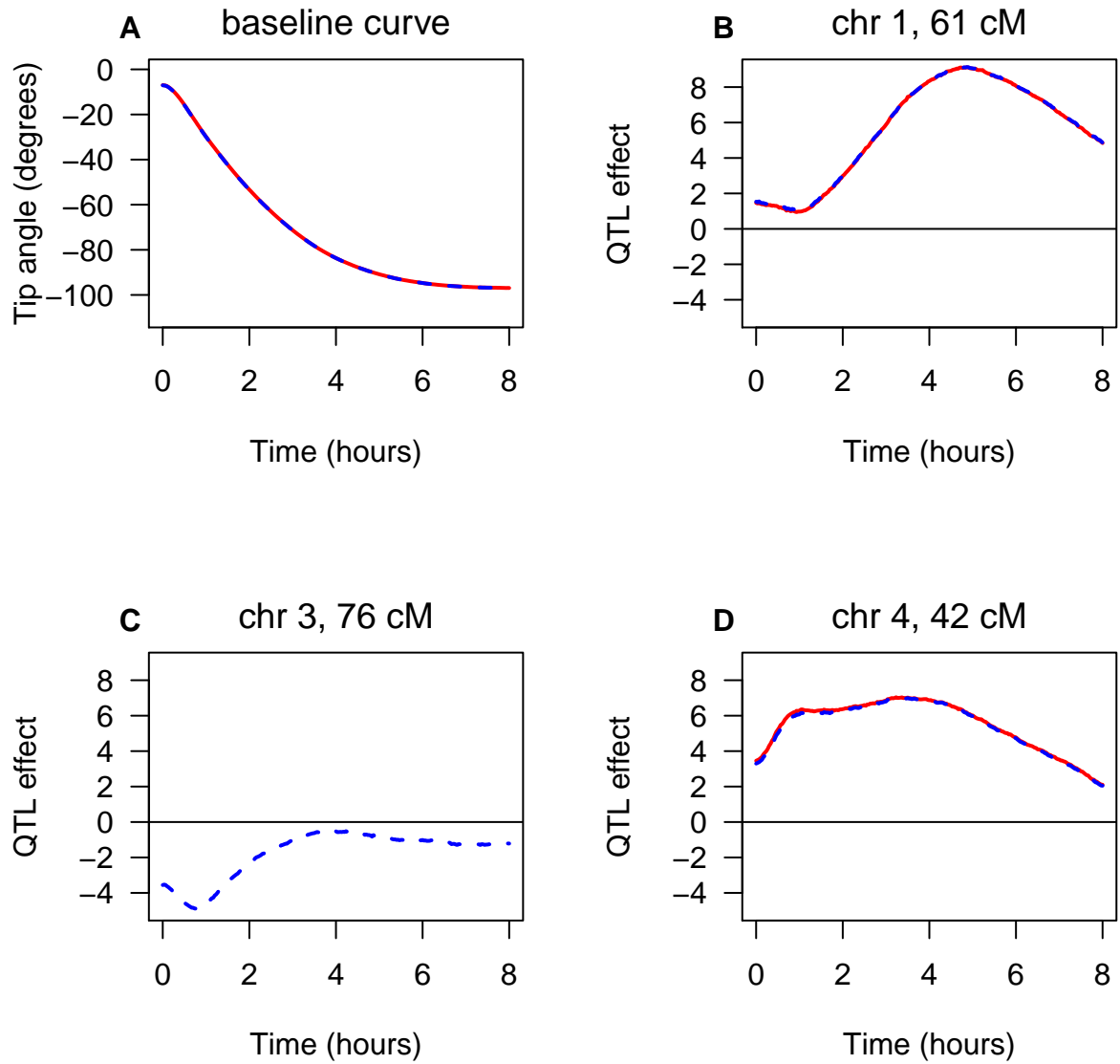}
\vspace{1cm}
\caption{\label{fp1} The regression coefficients estimated for
    the root tip angle data set. The red curves are for the two-QTL
    model (from the penalized-SLOD criterion) and the blue dashed
    curves are for the three-QTL model (from the penalized-MLOD
    criterion). Positive values for the QTL effects indicate that the
    Cvi allele increases the tip angle phenotype.}
\end{center}
\end{figure}

To further characterize the effects of the QTL in the context of the
inferred multiple-QTL models, we fit the selected multiple-QTL models
at each time point, individually. For the models derived by the
penalized-SLOD and penalized-MLOD criteria, the estimated baseline
function and the estimated QTL effects, as a function of time, are
shown in Figure~\ref{fp1}. The estimated QTL effects in panels B--D
are for the difference between the Cvi allele and the Ler allele.

The effects of the QTL on chromosomes 1 and 4 are approximately the
same, whether or not the chromosome 3 QTL is included in the model.
The chromosome 1 QTL has greatest effect at later time points, while
the chromosome 4 QTL has greatest effect earlier and over a wider
interval of time.  For both QTL, the Cvi allele increases the root tip
angle phenotype.
The chromosome 3 QTL, identified only with the penalized-MLOD
criterion, has an effect at early time points, and only for a brief
interval of time, and for this QTL, the Ler allele increases the root tip
angle phenotype.

\clearpage

\centerline{\sffamily \textbf{Simulations}}

In order to investigate the performance of our proposed approaches and
compare them to existing methods, we performed several computer
simulation studies. While numerous methods for QTL mapping with
function-valued traits have been described, we were unsuccessful,
despite considerable effort, to employ the software for
\citet{Yang2009}, \citet{Yap2009}, \citet{Min2011}, or
\citet{Sillanpaa2012}. Thus our main focus for comparison was to the
estimating equation approach of \citet{Xiong2011}. This method has
been implemented only for a single-QTL genome scan, and so we compare
our approach to that method in the presence of a single QTL.
In these
single-QTL models, we also considered a simple parametric approach:
fit growth curves for each individual \citep{Kahm2010} and then apply
multivariate QTL analysis \citep{Knott2000} with the estimated
parameters as phenotypes. In the context of multiple-QTL models, we
considered only the two variants of our own approach, the
penalized-SLOD and penalized-MLOD criteria.

\textbf{Single-QTL models:}
To compare our approach to that of \citet{Xiong2011}, and to a simple
parametric approach, in the context of
a single-QTL model, we considered the simulation setting described in
\citet{Yap2009}, though exploring a range of QTL effects.

We simulated an intercross with sample sizes of 100, 200, or 400, and a
single chromosome of length 100~cM,
with 6 equally spaced markers and with a QTL at 32~cM. The associated
phenotypes was sampled from a multivariate normal distribution with
the mean curve following a logistic function, $g(t) = \frac{a}{1+b e^{-rt}}$.
The AA genotype had $a=29, b=7, r=0.7$; the AB genotype had $a=28.5,
b=6.5, r=0.73$; and the BB genotype had $a=27.5,
b=5, r=0.75$. The shape of growth curve with this parameter shown in
Figure~S3. Each individual is observed at 10 time
points.

The residual error was assumed to be multivariate normal with a
covariance structure $c\Sigma$. The constant $c$ controls the overall
error variance, and $\Sigma$ was chosen to have one of three different
covariance structures: (1) auto-regressive with $\sigma^2 =3, \rho =
0.6$, (2) equi-correlated with $\sigma^2 =3, \rho = 0.5$, or (3) an
``unstructured'' covariance matrix, as given in \citet{Yap2009} (shown
in Table~S2).

The parameter $c$ was given a range of values, which define the
percent phenotypic variance explained by the QTL (the heritability).
The effect of the QTL varies with time; we took the mean heritability
across time as an overall summary.  For the auto-regressive and
equi-correlated covariance structures, we used $c = 1,2,3,6$; for the
unstructured covariance matrix, we took $c = 0.5, 1, 2, 3$.
The
heritabilities, as a function of time, for each covariance structure
and for each value of the parameter $c$, are shown in
Figure~S4.

For each of 10,000 simulation replicates, we applied our SLOD and MLOD
methods, using Haley-Knott regression \citep{Haley1992}, and the two
versions of the method of \citet{Xiong2011}, EE(Wald) and
EE(Residual). We further applied a simple parametric approach: We fit
the logistic growth model to each individual's phenotype data using
the R package grofit \citep{Kahm2010}, and then used the estimated
model parameters as phenotypes, applying the multivariate QTL mapping
method of \citet{Knott2000}. For all five approaches, we fit a
three-parameter QTL model (that is, allowing for dominance).

\begin{figure}[!ht]
\begin{center}
\includegraphics{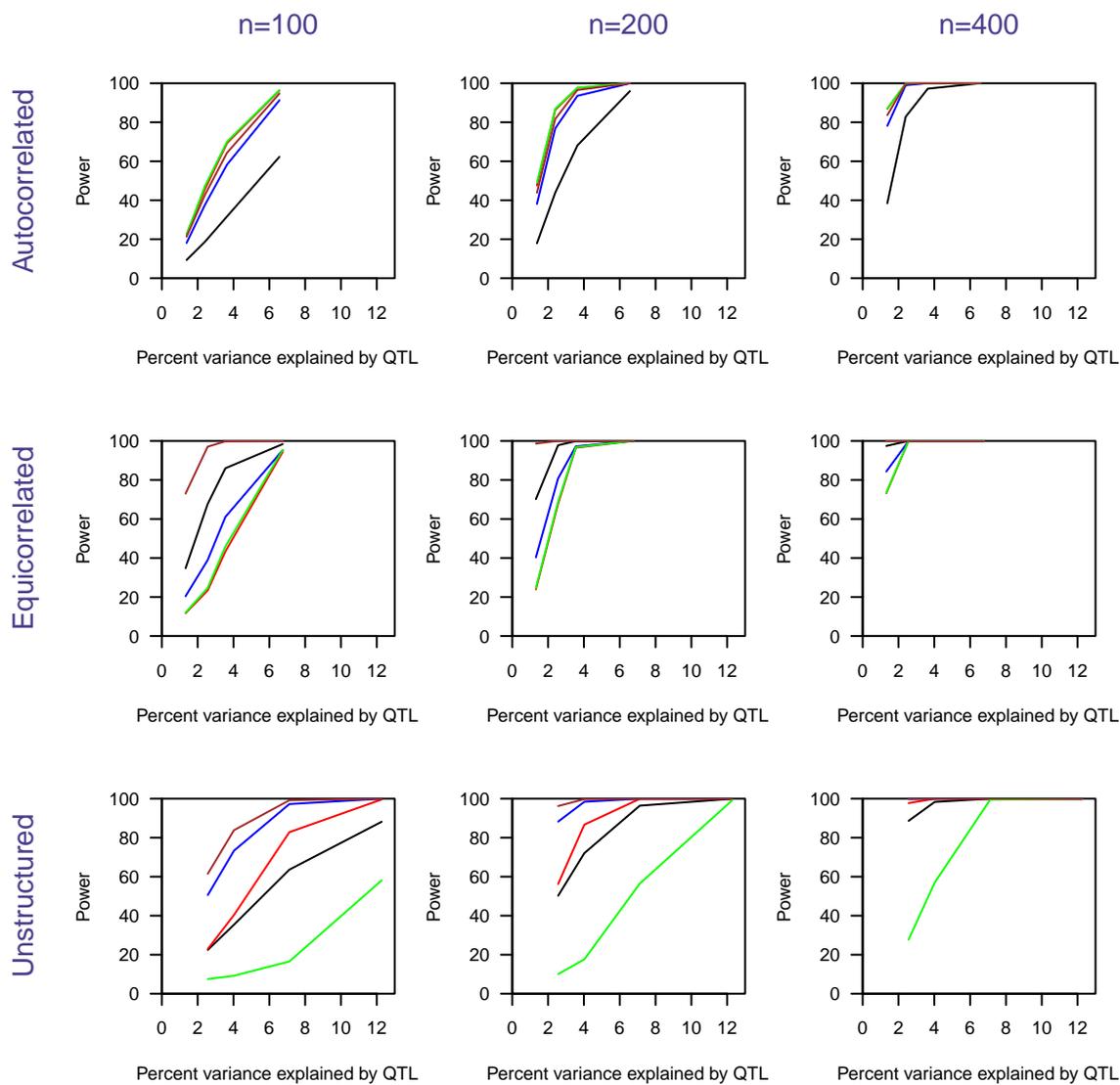}
\vspace{1cm}
\caption{\label{power1} Power as a function of the percent phenotypic
  variance explained by a single QTL. The first column is for $n=100$, the second column is for $n=200$ and the third column is for $n=400$. The three rows correspond to the
  covariance structure (autocorrelated, equicorrelated, and
  unstructured).  In each panel, SLOD is in red, MLOD is in blue,
  EE(Wald) is in brown, EE(Residual) is in green, and
  parametric is in black.}
\end{center}
\end{figure}

The estimated power to detect the QTL as a function of heritability
due to the QTL, for $n=100, 200, 400$ and for the three different covariance
structures, is shown in Figure~\ref{power1}.
With the autocorrelated variance structure, all methods other than
parametric approach gave similar power. With the equicorrelated
variance structure, EE(Wald) had higher power than the other four
methods, and the parametric approach was second-best.
In the unstructured variance setting, EE(Wald) and MLOD method worked better
than the other three methods. EE(Residual) didn't work well in
this setting.

The precision of QTL mapping, measured by the root mean square
error in the estimated QTL position, is displayed in Figure~S5.
Performance, in terms of precision, corresponds quite closely to the
performance in terms of power: when power is high, the RMS error of
the estimated QTL position is low, and vice versa.

A possible weakness of the SLOD and MLOD approaches, in not making
use of the function-valued form of the phenotypes, is that the methods
may suffer lower power in the case of noisy phenotypes. To investigate
this possibility, we repeated the simulations with $n=200$, adding
independent, normally distributed errors (with standard deviation 1 or
2) at each time point.

\begin{figure}[!ht]
\begin{center}
\includegraphics{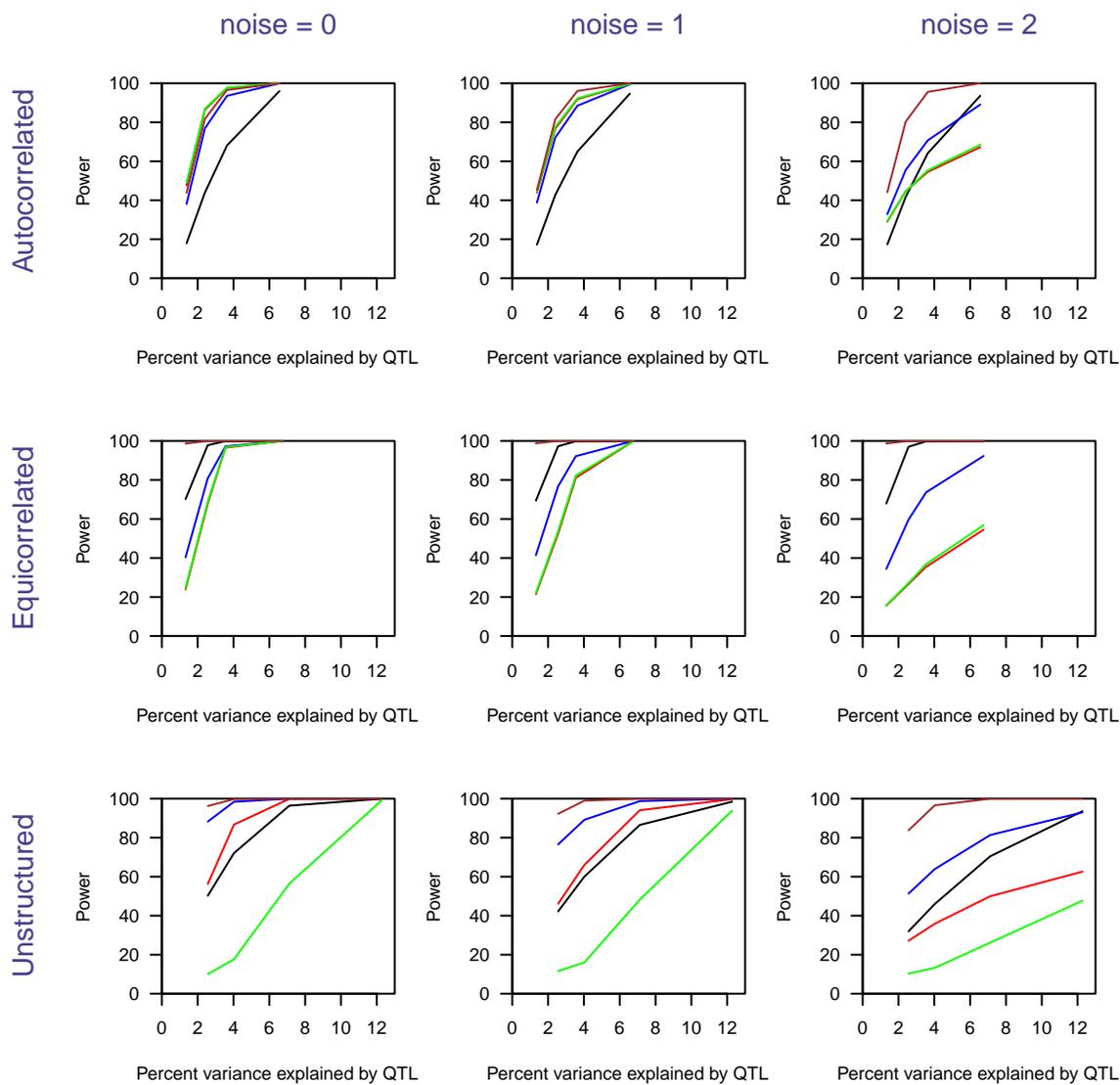}
\vspace{1cm}
\caption{\label{power2}
  Power as a function of the percent phenotypic variance explained by
  a single QTL, with additional noise added to the phenotypes.  The
  first column has no additional noise; the second and third columns
  have independent normally distributed noise added at each time
  point, with standard deviation 1 and 2, respectively.  The three
  rows correspond to the covariance structure (autocorrelated,
  equicorrelated, and unstructured).  In each panel, SLOD is in red,
  MLOD is in blue, EE(Wald) is in brown, EE(Residual) is in green, and
  parametric is in black. The percent variance explained by the QTL
  on the x-axis refers, in each case, to the variance explained in the
  case of no added noise.}
\end{center}
\end{figure}

The estimated power to detect the QTL as a function of heritability
due to the QTL, for added noise with SD $=0, 1, 2$ and the three
different covariance structures, is shown in Figure~\ref{power2}.
The power of the SLOD, MLOD, the EE(Residual) methods are greatly
affected by the introduction of noise. EE(Wald) and the parametric
methods are relatively robust to the introduction of noise.
Overall, the EE(Wald) method continued to perform best.

For smooth traits with autocorrelated errors, the SLOD and MLOD
methods works similarly to EE(Wald) and EE(Residual). However, if we
have large measurement error or have different variance structure, the
EE(Wald) method is a robust choice. The parametric approach was more
affected by the nature of the residual variance structure than by the
addition of random noise.

In terms of computation time, in this simulation study, the MLOD and
SLOD methods were about 3 times faster than EE(Residual), and they
were about 265 times faster than the EE(Wald) method, with five basis
functions used in the latter.

\textbf{Multiple-QTL models:}
To investigate the performance of the penalized-SLOD and
penalized-MLOD criteria in the context of multiple QTL,
we simulated data from a three-QTL model modeled after that estimated
from the root tip angle data of \citet{Moore2013}, considered in the
Application section.

We assumed that the mean curve for the root tip angle phenotype
followed a cubic polynomial, $y = a + b t + c t^2 + d t^3$,
and assumed that the effect of each QTL also followed such a cubic
polynomial. Fitting this parametric model with the three QTL derived
by the penalized-MLOD criterion, we obtained the following estimates.
The parameters of the baseline were
$(a,b,c,d) = (-0.238, -265.248, 229.405, -59.771)$.
The QTL effect for the QTL chromosome 1 at 61~cM
had parameters $(0.209, 8.729, 1.602, -9.054)$.
A second QTL, on chromosome 3 at 76~cM, had parameters
$(-1.887, 3.414, -4.220, 2.265)$. The third QTL, on chromosome 4 at
40~cM, had parameters $(2.003, 11.907, -28.647, 15.311)$.
The baseline function and the QTL effect curves are shown in Figure~S5.

The four parameters for a given individual were drawn from a
multivariate normal distribution with mean defined by the QTL
genotypes and variance matrix estimated from the root
tip angle data as:
\begin{equation*}
  \Sigma = \begin{pmatrix}
 58.99 &  -177.77 &  185.11 &  -45.44 \\
 -177.77 & 3848.70 & -7274.83 & 3595.37 \\
   185.11& -7274.83& 16897.56 &-9702.32\\
   -45.44&  3595.37& -9702.32 & 6096.71 \\
  \end{pmatrix}
\end{equation*}

In addition, normally distributed measurement error (with mean 0 and
variance 1) was added to the phenotype at each time point for each
individual. Phenotypes are taken at 241 equally spaced time points in
the interval of 0 to 1. We considered two sample sizes: $n=162$ (as in
the \citet{Moore2013} data) and twice that, $n=324$.

We performed 2000 simulation replicates. For each replicate, we
applied a stepwise model selection approach with each of the
penalized-SLOD and penalized-MLOD criteria.
The simulation results are shown in Table~\ref{table1}.

The penalized-SLOD criterion had higher power to detect the first and
third QTL, while the penalized-MLOD criterion had higher power to
detect the second QTL.
With the larger sample size, the power to detect QTL increased,
and the standard error of the estimated QTL position decreased.

The estimated false positive rates at $n=162$ were 3.3 and 0.9\% for
the penalized-SLOD and penalized-MLOD criteria, respectively. At the
larger sample size, $n=324$, the corresponding false positive rates
were 4.1 and 0.8\%.

\clearpage
\centerline{\sffamily \textbf{Discussion}}

Automated phenotype measurement is an accelerating trend across
biological scales, from microorganisms to crop plants. This push
for increasing automation makes it feasible to increase the
dimensionality of phenotype data sets, for example by adding
time. The trend toward higher-dimensional phenotype data sets from
genetically structured populations has created a need for new
statistical genetic methods, and computational speed can be an
important factor in the application of such methods.

Methods for the genetic analysis of function-valued phenotypes have
mostly focused on single-QTL models \citep{Ma2002, Yang2009, Yap2009,
Xiong2011}. Bayesian multiple-QTL methods, using Markov chain Monte
Carlo, have also been proposed \citep{Min2011, Sillanpaa2012}, but
they can be computationally intensive and not easily implemented. We
propose two simple LOD-type statistics that integrate the information
across time points and extend them, using the approach of
\citet{Broman2002}, for multiple-QTL model selection.

The basis of our approach is the analysis of each time point
individually. This works well when the function-valued trait is
smooth, as in the data from \citet{Moore2013}, and has the benefit of
providing results that are easily interpreted, such as the QTL effects
displayed in Figure~\ref{fp1}. With unequally-spaced time points or
appreciable missing data, the approach may require some modification,
such as first performing some interpolation or smoothing.
The
performance of our approaches deteriorated with added noise, but again
this may be at least partly alleviated by pre-smoothing.
An important advantage
of our approach is the ability to incorporate information from
multiple QTL in the analysis of function-valued phenotypes, which
should improve power and lead to better separation of linked QTL.

A weakness of our approach is that it largely ignores the correlations
across time. \citet{Ma2002} and \citet{Yang2009} paid careful attention
to this aspect, using an autoregressive model for the residual
variance matrix. Our current neglect of this aspect may result in
loss of efficiency, particularly in the estimates of the QTL
effects. However by ignoring this assumption we gain much speed, and
our simulation studies indicate that the approach exhibits reasonable
power to detect QTL in many situations. The EE(Wald)
method of \citet{Xiong2011} showed the best performance among all
methods considered, though at the expense of considerably greater
computation time.

\citet{Manichaikul2009} extended the work of \citet{Broman2002} by
considering pairwise interactions among QTL. Our approach may be
similarly extended to consider interactions.

An alternative approach to the QTL analysis of function-valued traits
is to first fit a parametric model to each individual's curve and then
treat the estimated parameters from such a model as
phenotypes. (The method exhibited less-than-ideal performance in our
simulation study, likely due to poor model fit with the simulated error
structures.)
Multiple-QTL mapping methods could readily be applied to
each such parameter, individually. The advantage of our approach, to
consider each time point individually, is in the simpler interpretation
of the results.

Software implementing our methods have been implemented as a package
for R \citep{R}, funqtl
({\small \verb|https://github.com/ikwak2/funqtl|}).

\clearpage
\centerline{\sffamily \textbf{Acknowledgments}}
The authors thank \'Saunak Sen and two anonymous
reviewers for
comments to improve the manuscript.
This work was supported in part by grant IOS-1031416 from the National
Science Foundation Plant Genome Research Program to E.P.S. and by
National Institutes of Health grant R01GM074244 to K.W.B.

\clearpage
\bibliographystyle{genetics}
\renewcommand*{\refname}{\centerline{\sffamily \normalsize \textbf{Literature Cited}}}
\bibliography{funqtl}

\begin{thebibliography}{17}
\expandafter\ifx\csname natexlab\endcsname\relax\def\natexlab#1{#1}\fi

\bibitem[{{\rm Broman} and {\rm Speed}(2002)}]{Broman2002}
{\rm Broman, K.~W.},  and {\rm T.~P. Speed}, 2002 A model selection approach
  for the identification of quantitative trait loci in experimental crosses.
\newblock J. R. Statist. Soc. B {\bf 64}: 641--656.

\bibitem[{{\rm Churchill} and {\rm Doerge}(1994)}]{Churchill1994}
{\rm Churchill, G.~A.},  and {\rm R.~W. Doerge}, 1994 Empirical threshold
  values for quantitative trait mapping.
\newblock Genetics {\bf 138}: 963--971.

\bibitem[{{\rm Haley} and {\rm Knott}(1992)}]{Haley1992}
{\rm Haley, C.~S.},  and {\rm S.~A. Knott}, 1992 A simple regression method for
  mapping quantitative trait loci in line crosses using flanking markers.
\newblock Heredity {\bf 69}: 315--324.

\bibitem[{{\rm Kahm} {\em et~al.\/}(2010){\rm Kahm}, {\rm Hasenbrink}, {\rm
  Lichtenberg-Frat\'e}, {\rm Ludwig} and {\rm Kschischo}}]{Kahm2010}
{\rm Kahm, M.}, {\rm G.~Hasenbrink}, {\rm H.~Lichtenberg-Frat\'e}, {\rm
  J.~Ludwig},  and {\rm M.~Kschischo}, 2010 {grofit}: Fitting biological growth
  curves with {R}.
\newblock J. Statist. Softw. {\bf 33}: 1--21.

\bibitem[{{\rm Kendziorski} {\em et~al.\/}(2002){\rm Kendziorski}, {\rm
  Cowley}, {\rm Greene}, {\rm Salgado}, {\rm Jacob} {\em
  et~al.\/}}]{Kendziorski2002}
{\rm Kendziorski, C.~M.}, {\rm A.~W. Cowley}, {\rm A.~S. Greene}, {\rm H.~C.
  Salgado}, {\rm H.~J. Jacob} {\em et~al.\/}, 2002 Mapping baroreceptor
  function to genome: a mathematical modeling approach.
\newblock Genetics {\bf 160}: 1687--1695.

\bibitem[{{\rm Knott} and {\rm Haley}(2000)}]{Knott2000}
{\rm Knott, S.~A.},  and {\rm C.~S. Haley}, 2000 Multitrait least squares for
  quantitative trait loci detection.
\newblock Genetics {\bf 156}: 899--911.

\bibitem[{{\rm Lander} and {\rm Botstein}(1989)}]{Lander1989}
{\rm Lander, E.~S.},  and {\rm D.~Botstein}, 1989 Mapping {M}endelian factors
  underlying quantitative traits using {RFLP} linkage maps.
\newblock Genetics {\bf 121}: 185--199.

\bibitem[{{\rm Ma} {\em et~al.\/}(2002){\rm Ma}, {\rm Casella} and {\rm
  Wu}}]{Ma2002}
{\rm Ma, C.}, {\rm G.~Casella},  and {\rm R.~L. Wu}, 2002 Functional mapping of
  quantitative trait loci underlying the character process: A theoretical
  framework.
\newblock Genetics {\bf 161}: 1751--1762.

\bibitem[{{\rm Manichaikul} {\em et~al.\/}(2009){\rm Manichaikul}, {\rm Moon},
  {\rm Sen}, {\rm Yandell} and {\rm Broman}}]{Manichaikul2009}
{\rm Manichaikul, A.}, {\rm J.~Y. Moon}, {\rm S.~Sen}, {\rm B.~S. Yandell},
  and {\rm K.~W. Broman}, 2009 A model selection approach for the
  identification of quantitative trait loci in experimental crosses, allowing
  epistasis.
\newblock Genetics {\bf 181}: 1077--1086.

\bibitem[{{\rm Min} {\em et~al.\/}(2011){\rm Min}, {\rm Yang}, {\rm Wang} and
  {\rm Wang}}]{Min2011}
{\rm Min, L.}, {\rm R.~Yang}, {\rm X.~Wang},  and {\rm B.~Wang}, 2011 Bayesian
  analysis for genetic architecture of dynamic traits.
\newblock Heredity {\bf 106}: 124--133.

\bibitem[{{\rm Moore} {\em et~al.\/}(2013){\rm Moore}, {\rm Johnson}, {\rm
  Kwak}, {\rm Livny}, {\rm Broman} {\em et~al.\/}}]{Moore2013}
{\rm Moore, C.~R.}, {\rm L.~S. Johnson}, {\rm I.-Y. Kwak}, {\rm M.~Livny}, {\rm
  K.~W. Broman} {\em et~al.\/}, 2013 High-throughput computer vision introduces
  the time axis to a quantitative trait map of a plant growth response.
\newblock Genetics {\bf 195}: 1077--1086.

\bibitem[{{\rm {R Core Team}}(2013)}]{R}
{\rm {R Core Team}}, 2013 {\em R: A Language and Environment for Statistical
  Computing\/}.
\newblock R Foundation for Statistical Computing, Vienna, Austria.

\bibitem[{{\rm Sillanp\"a\"a} {\em et~al.\/}(2012){\rm Sillanp\"a\"a}, {\rm
  Pikkuhookana}, {\rm Abrahamsson}, {\rm Knurr}, {\rm Fries} {\em
  et~al.\/}}]{Sillanpaa2012}
{\rm Sillanp\"a\"a, M.~J.}, {\rm P.~Pikkuhookana}, {\rm S.~Abrahamsson}, {\rm
  T.~Knurr}, {\rm A.~Fries} {\em et~al.\/}, 2012 Simultaneous estimation of
  multiple quantitative trait loci and growth curve parameters through
  hierarchical bayesian modeling.
\newblock Heredity {\bf 108}: 134--146.

\bibitem[{{\rm Xiong} {\em et~al.\/}(2011){\rm Xiong}, {\rm Goulding}, {\rm
  Carlson}, {\rm Tecott}, {\rm McCulloch} {\em et~al.\/}}]{Xiong2011}
{\rm Xiong, H.}, {\rm E.~H. Goulding}, {\rm E.~J. Carlson}, {\rm L.~H. Tecott},
  {\rm C.~E. McCulloch} {\em et~al.\/}, 2011 A flexible estimating equations
  approach for mapping function-valued traits.
\newblock Genetics {\bf 189}: 305--316.

\bibitem[{{\rm Yang} {\em et~al.\/}(2009){\rm Yang}, {\rm Wu} and {\rm
  Casella}}]{Yang2009}
{\rm Yang, J.}, {\rm R.~L. Wu},  and {\rm G.~Casella}, 2009 Nonparametric
  functional mapping of quantitative trait loci.
\newblock Biometrics {\bf 65}: 30--39.

\bibitem[{{\rm Yap} {\em et~al.\/}(2009){\rm Yap}, {\rm Fan} and {\rm
  Wu}}]{Yap2009}
{\rm Yap, J.~S.}, {\rm J.~Fan},  and {\rm R.~Wu}, 2009 Nonparametric modeling
  of longitudinal covariance structure in functional mapping of quantitative
  trait loci.
\newblock Biometrics {\bf 65}: 1068--1077.

\bibitem[{{\rm Zeng} {\em et~al.\/}(2000){\rm Zeng}, {\rm Liu}, {\rm Stam},
  {\rm Kao}, {\rm Mercer} {\em et~al.\/}}]{Zeng2000}
{\rm Zeng, Z.~B.}, {\rm J.~J. Liu}, {\rm L.~F. Stam}, {\rm C.~H. Kao}, {\rm
  J.~M. Mercer} {\em et~al.\/}, 2000 Genetic architecture of a morphological
  shape difference between two {D}rosophila species.
\newblock Genetics {\bf 154}: 299--310.

\end{thebibliography}

\newpage

\renewcommand{\arraystretch}{2}
\begin{table}[!ht]
  \caption{Simulation results for the SLOD and MLOD criteria, for a
    three-QTL model modeled after the \citet{Moore2013} data.\label{table1}}
  \begin{center}
    \begin{tabular}{lcccccc}
      \hline
      & & \multicolumn{2}{c}{Mean (SE) estimated location} & &
      \multicolumn{2}{c}{Power} \\ \cline{3-4}\cline{6-7}
      & True location & SLOD & MLOD && SLOD & MLOD \\ \hline
      $n$=162         & 61 & 60.9 (6.1) & 61.1 (4.9) && 89 & 54 \\
                      & 76 & 65.0 (18.3) & 71.4 (11.5) && 12 & 15 \\
                      & 40 & 40.0 (4.9) & 40.4 (3.4) && 82 & 77 \\ \hline
      $n$=324         & 61 & 61.3 (2.6) & 61.3 (3.8) && 100 & 59 \\
                      & 76 & 72.0 (9.7) & 74.5 (4.8) && 31 & 43 \\
                      & 40 & 40.0 (2.1) & 40.1 (2.0) && 100 & 91 \\ \hline
    \end{tabular}
  \end{center}

\bigskip

Note: Locations are in cM.

\end{table}

\newpage

\centerline{\sffamily \textbf{Figure Legends}}

\begin{hanging}

\item \textbf{Figure 1.}
Signed LOD scores from single-QTL genome scans, with each time point
considered individually.

\item \textbf{Figure 2.}
The SLOD, MLOD, EE(Wald) and EE(Residual) curves for the root tip
angle data. A red horizontal line indicates the calculated 5\%
permutation-based threshold.

\item \textbf{Figure 3.}
SLOD and MLOD profiles for a multiple-QTL model for the root tip angle data set.

\item \textbf{Figure 4.}
The regression coefficients estimated for the root tip angle data set. The
red curves are for the two-QTL model (from the penalized-SLOD
criterion) and the blue dashed curves are for the three-QTL model
(from the penalized-MLOD criterion).  Positive values for the QTL
effects indicate that the Cvi allele increases the tip angle
phenotype.

\item \textbf{Figure 5.}
Power as a function of the percent phenotypic
  variance explained by a single QTL. The first column is for $n=100$, the second column is for $n=200$ and the third column is for $n=400$. The three rows correspond to the
  covariance structure (autocorrelated, equicorrelated, and
  unstructured).  In each panel, SLOD is in red, MLOD is in blue,
  EE(Wald) is in brown, EE(Residual) is in green, and parametric is in
  black.

\item \textbf{Figure 6.}Power as a function of the percent phenotypic variance explained by
  a single QTL, with additional noise added to the phenotypes.  The
  first column has no additional noise; the second and third columns
  have independent normally distributed noise added at each time
  point, with standard deviation 1 and 2, respectively.  The three
  rows correspond to the covariance structure (autocorrelated,
  equicorrelated, and unstructured).  In each panel, SLOD is in red,
  MLOD is in blue, EE(Wald) is in brown, EE(Residual) is in green, and
  parametric is in black.  The percent variance explained by the QTL
  on the x-axis refers, in each case, to the variance explained in the
  case of no added noise.
\end{hanging}

\newpage

\newpage

\newpage

\newpage

\newpage

\newpage

\renewcommand{\thefigure}{\textbf{S\arabic{figure}}}
\renewcommand{\figurename}{\textbf{Figure}}

\renewcommand{\thetable}{\textbf{S\arabic{table}}}
\renewcommand{\tablename}{\textbf{Table}}

\vspace*{8mm}
\begin{center}

\textbf{\Large A simple regression-based method to map \\[6pt]
  quantitative trait loci underlying function-valued
  phenotypes}

\bigskip \bigskip \bigskip \bigskip

\textbf{\Large SUPPLEMENT}

\bigskip \bigskip
\bigskip \bigskip

{\large Il-Youp Kwak$^*$, Candace R. Moore$^\dagger$, Edgar
  P. Spalding$^\dagger$, Karl W. Broman$^{\ddagger}$}

\bigskip \bigskip

Departments of $^*$Statistics, $^\dagger$Botany, and $^\ddagger$Biostatistics and Medical
Informatics, \\
University of Wisconsin--Madison, Madison, Wisconsin 53706
\end{center}

\vfill

\hfill
{\footnotesize 15 May 2014}

\clearpage

\setcounter{figure}{0} 

\begin{figure}[p]
\centerline{\includegraphics{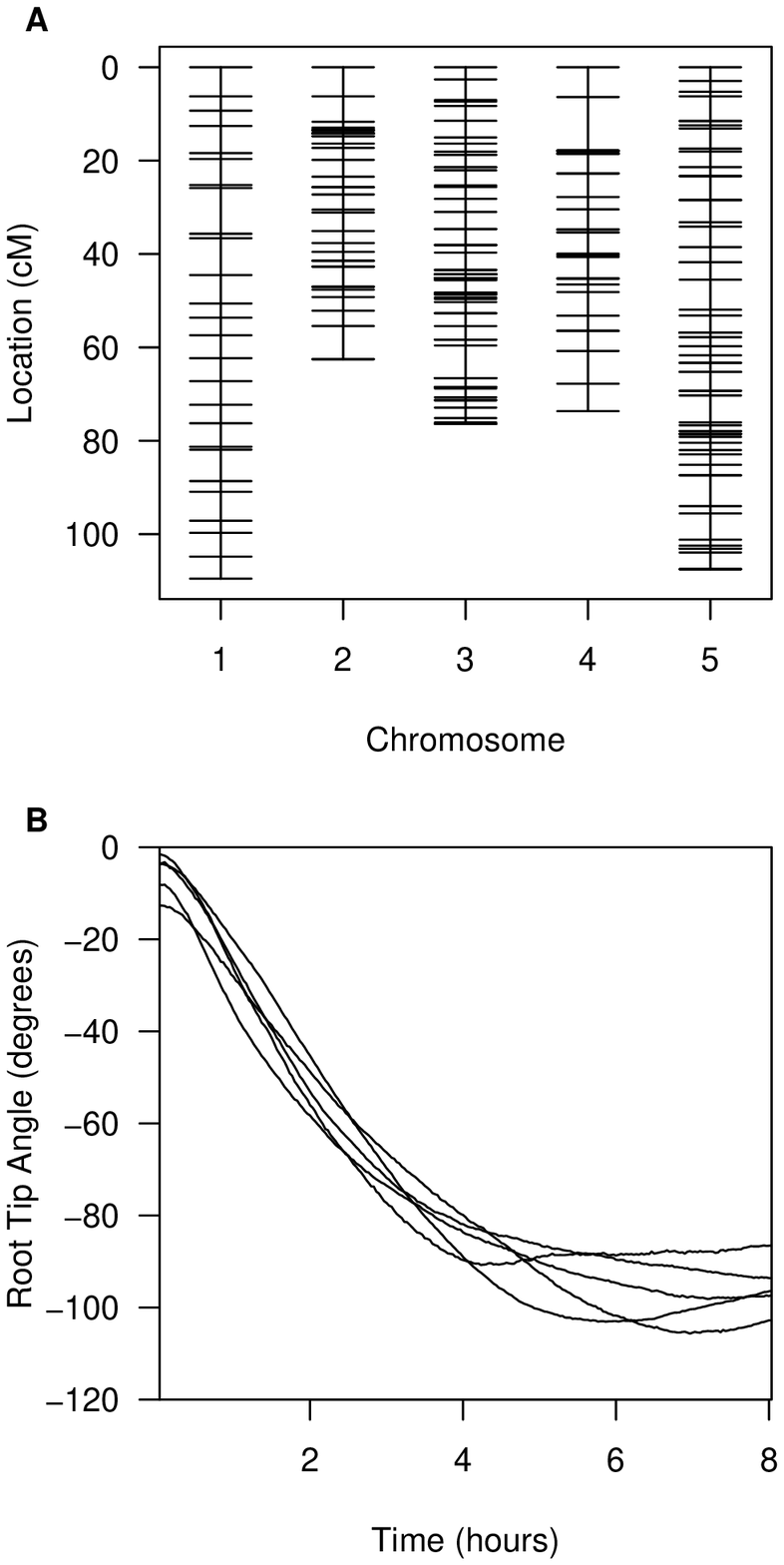}}

\caption{Genetic map of typed genetic markers (A)
     and function-valued phenotypes for five randomly selected
     Arabidopsis RIL (B), for data from Edgar Spalding and colleagues.}
\end{figure}

\clearpage

\begin{figure}[p]
\centerline{\includegraphics{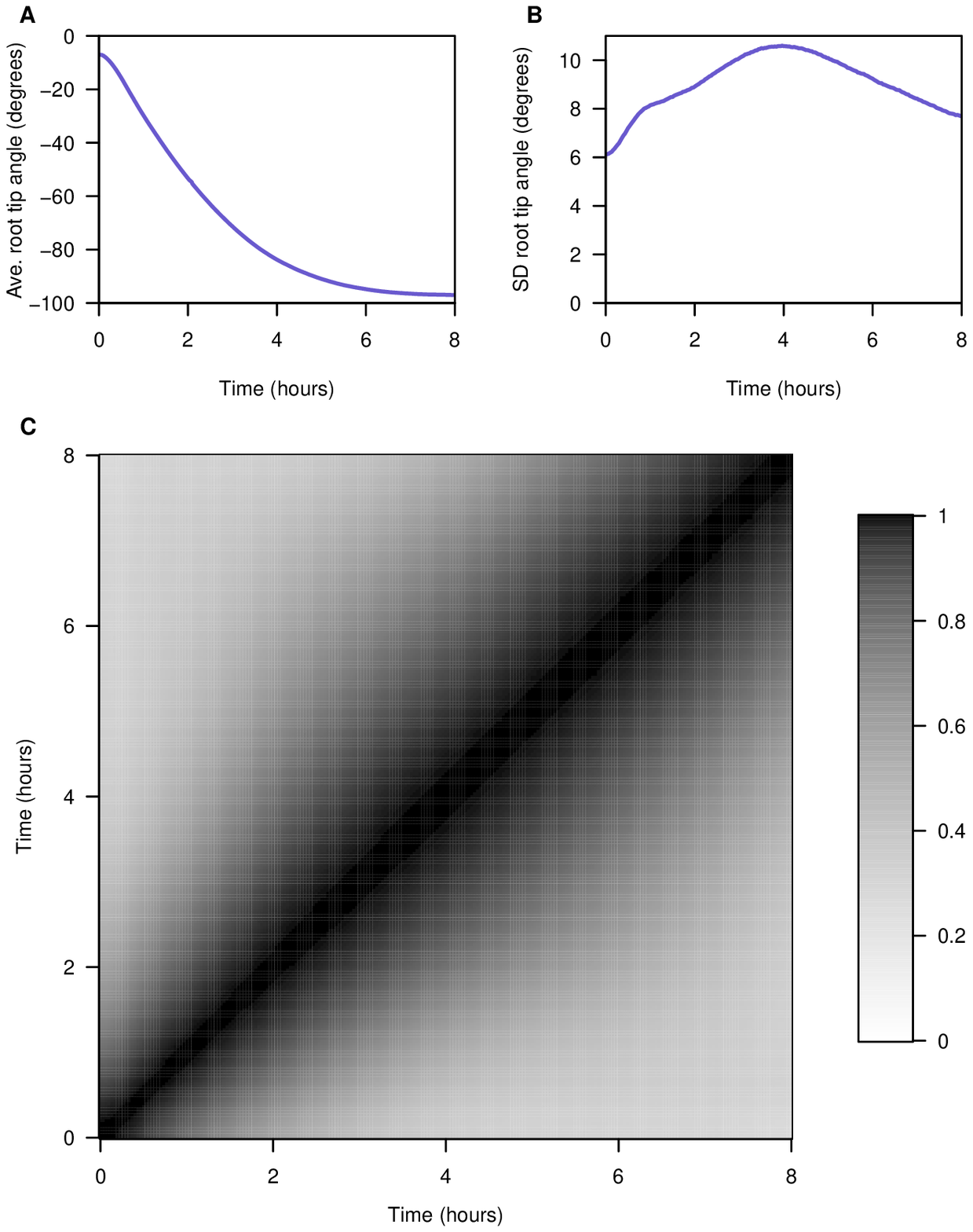}}
\vspace{1cm}

\caption{Average (A) and standard deviation (B) of the root tip angle phenotype at
  each individual time point, and the correlations between time points (C).}
\end{figure}

\clearpage

\begin{figure}[!ht]
\begin{center}
\includegraphics{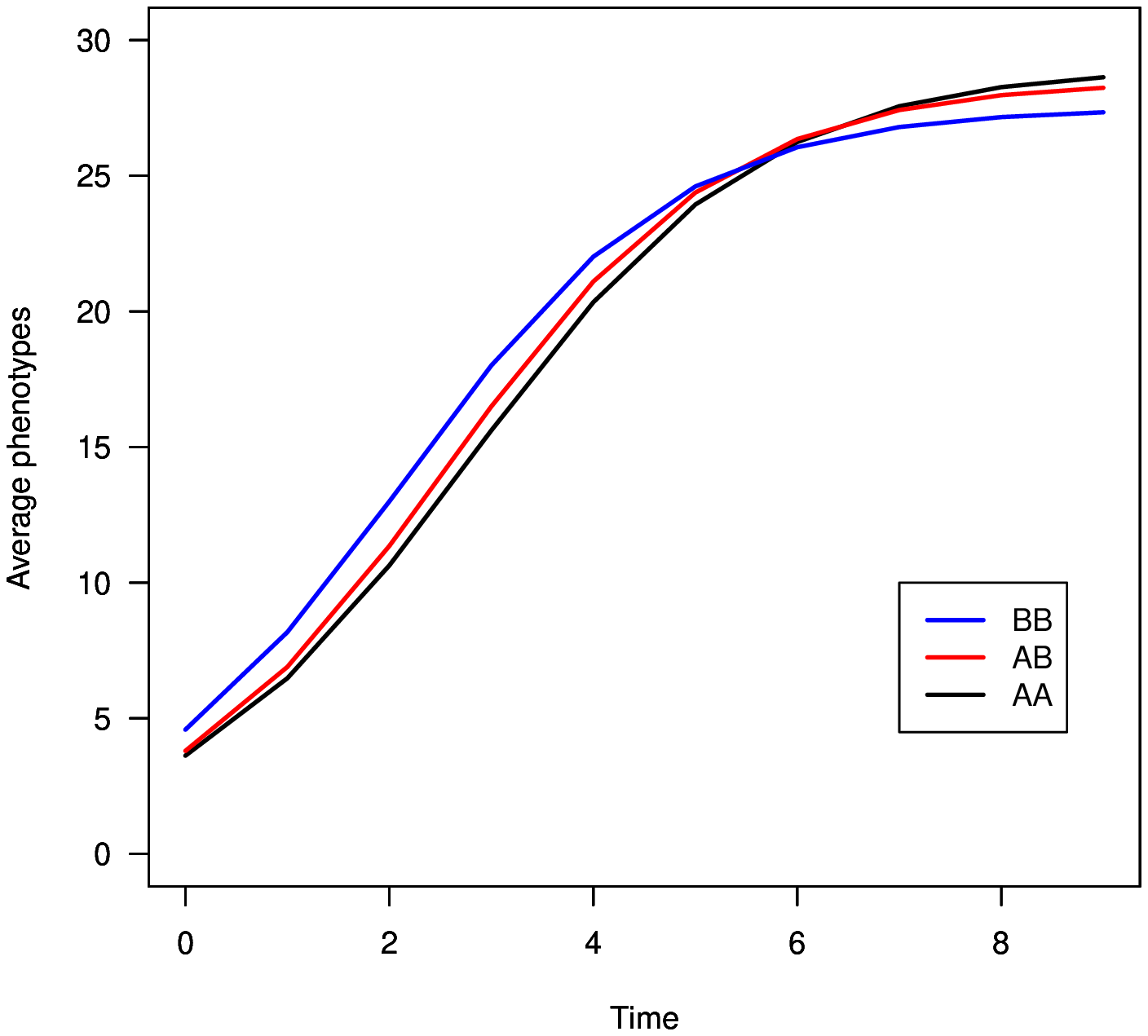}
\vspace{1cm}
 \caption{Growth curves for the three QTL genotypes in the single-QTL
   simulation study.}
\end{center}
\end{figure}

\clearpage

\begin{figure}[!ht]
\begin{center}
\includegraphics{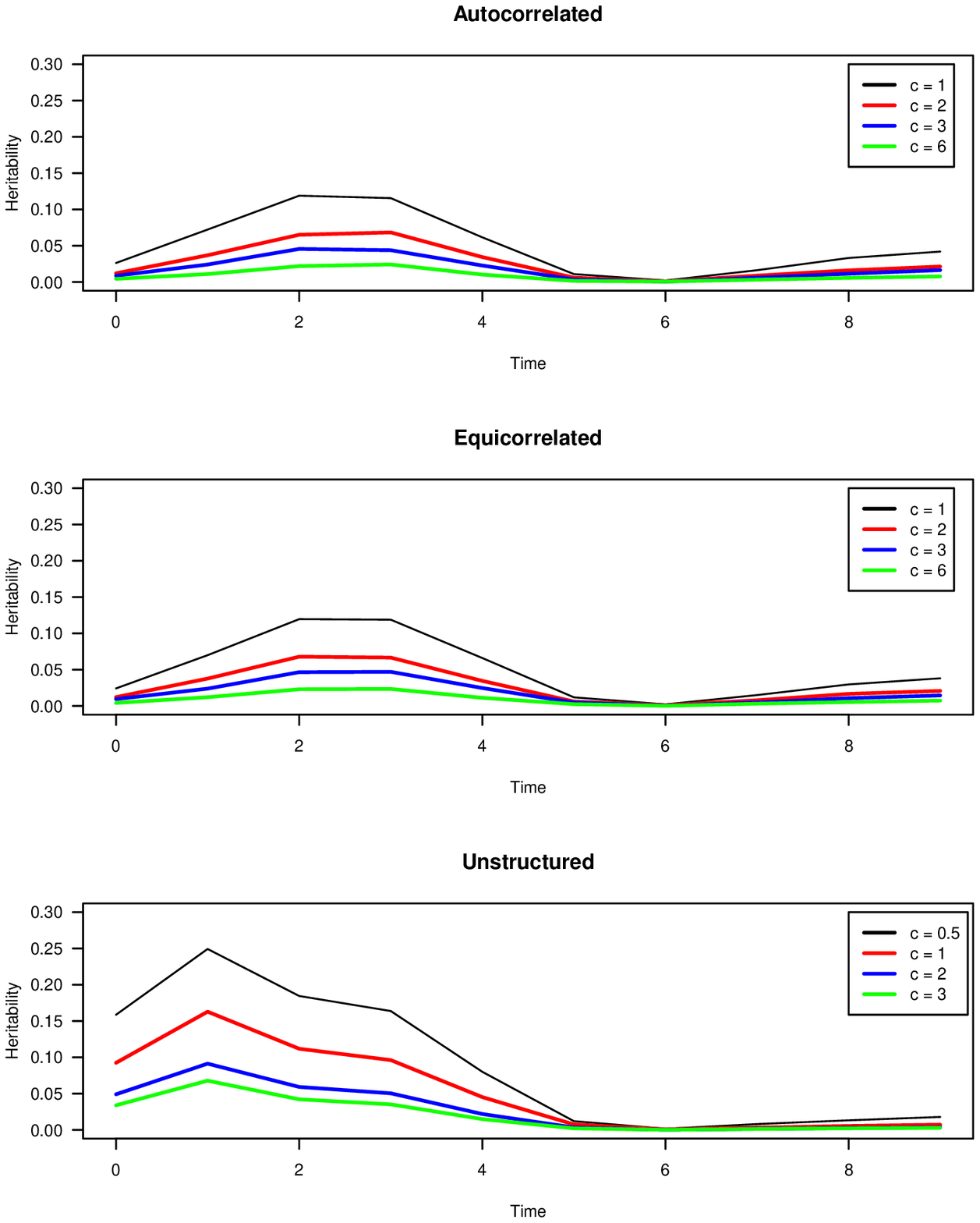}
\vspace{1cm}
 \caption{The heritability for each time point in the single-QTL
   simulation study, for the three assumed variance structures and the
   chosen values for the $c$ parameter.}
\end{center}
\end{figure}

\newpage

\begin{figure}[!ht]
\begin{center}
\includegraphics{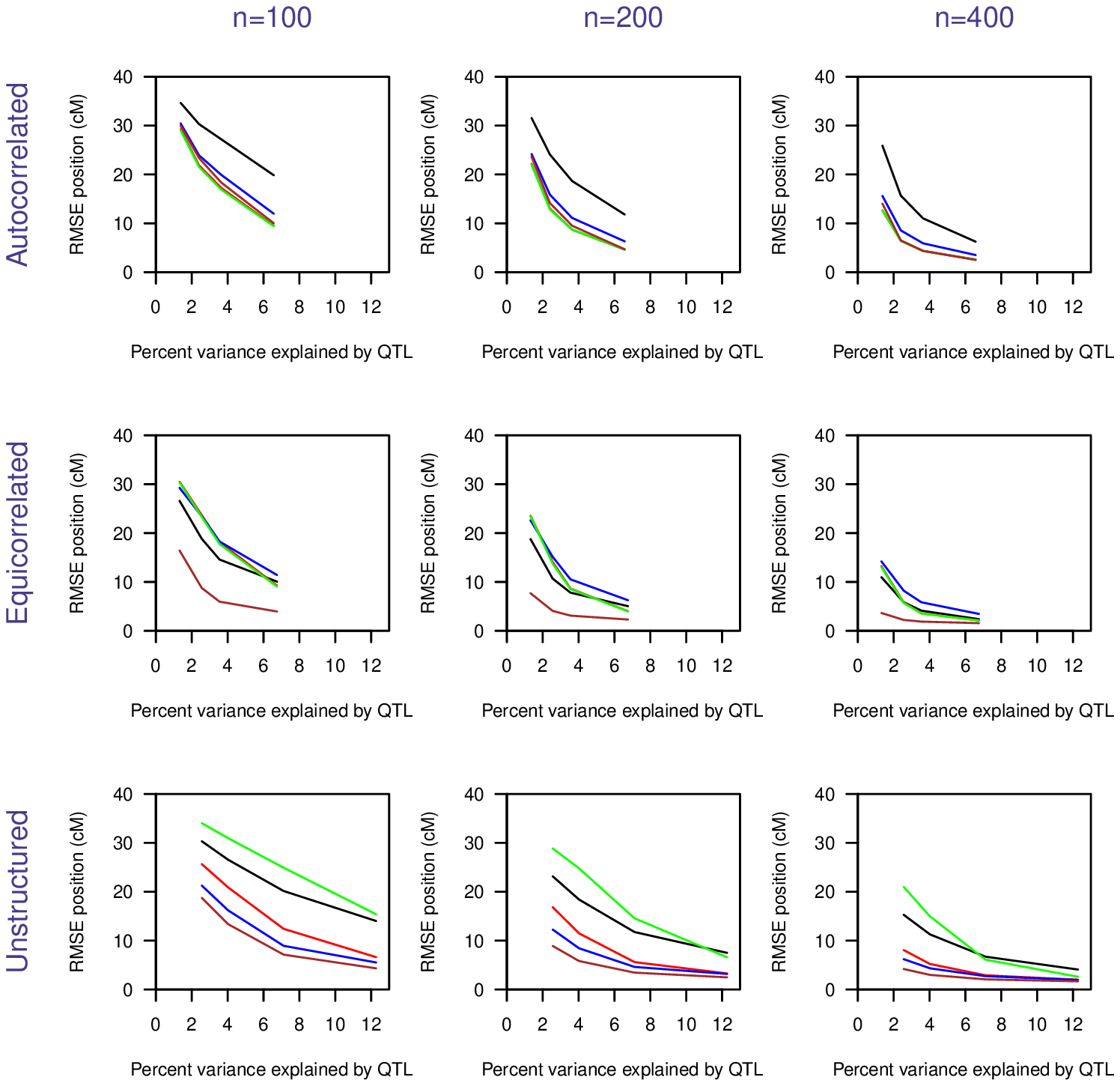}
\vspace{1cm}
 \caption{Root Mean Square Error (RMSE) of the estimated
   QTL position as a function of the percent variance explained by a
   single QTL.  The first column is for $n=100$, the second column is for $n=200$ and the third column is
   for $n=400$. The three rows correspond to the covariance structure
   (autocorrelated, equicorrelated, and unstructured).  In each panel,
   SLOD is in red, MLOD is in blue, EE(Wald) is in brown, EE(Residual)
   is in green, and parametric is in black.}
\end{center}
\end{figure}

\newpage

\begin{figure}[!ht]
\begin{center}
\includegraphics{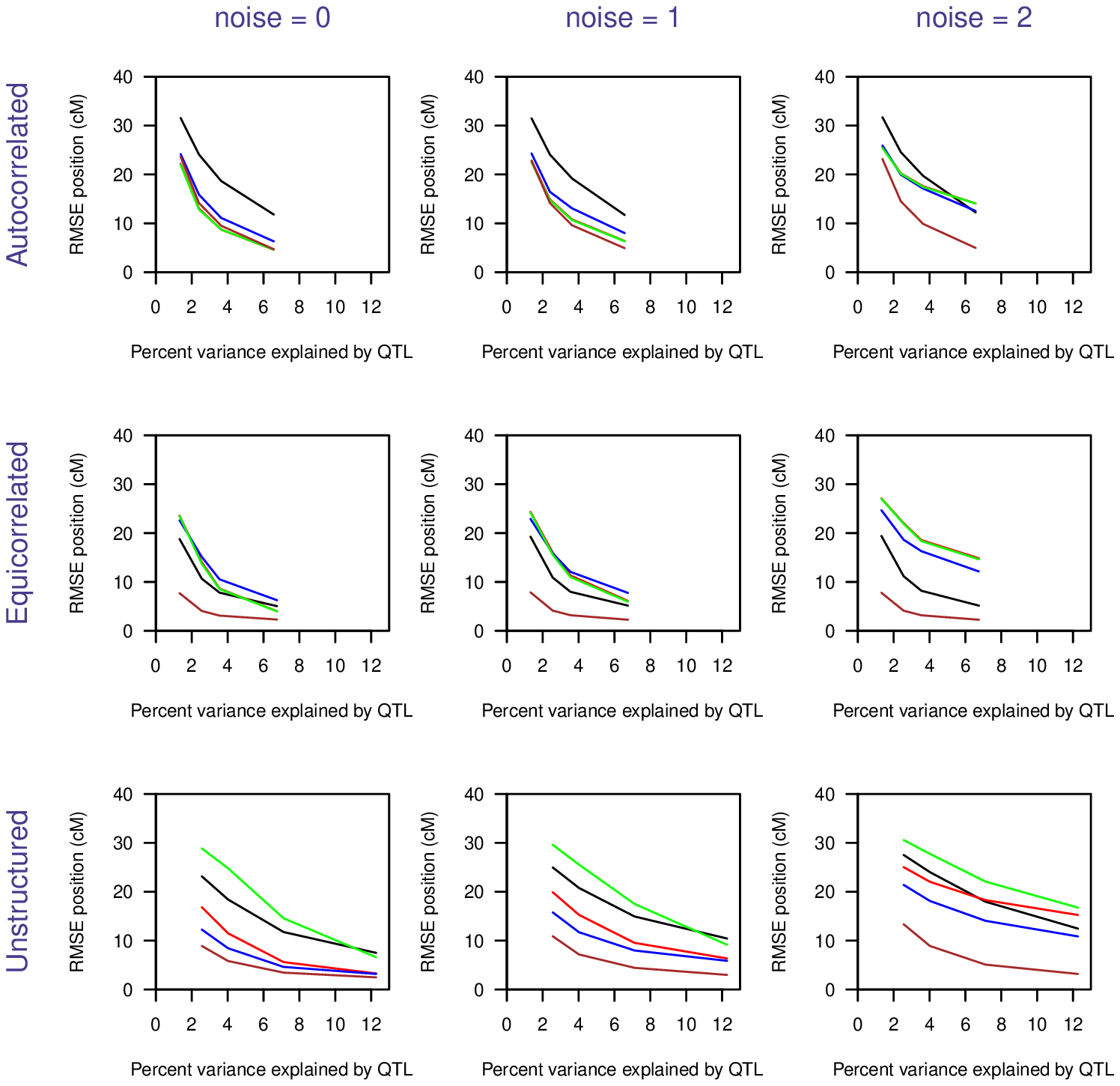}
\vspace{1cm}
 \caption{Root Mean Square Error (RMSE) of the estimated QTL position
   as a function of the percent variance explained by a single QTL,
   with additional noise added to the phenotypes.  The first column
   has no additional noise; the second and third columns have
   independent normally distributed noise added at each time point,
   with standard deviation 1 and 2, respectively. The three rows
   correspond to the covariance structure (autocorrelated,
   equicorrelated, and unstructured). In each panel, SLOD is in red,
   MLOD is in blue, EE(Wald) is in brown, EE(Residual) is in green,
   and parametric is in black. The percent variance explained by the
   QTL on the x-axis refers, in each case, to the variance explained
   in the case of no added noise.}
\end{center}
\end{figure}

\clearpage

\begin{figure}[!ht]
\begin{center}
\includegraphics{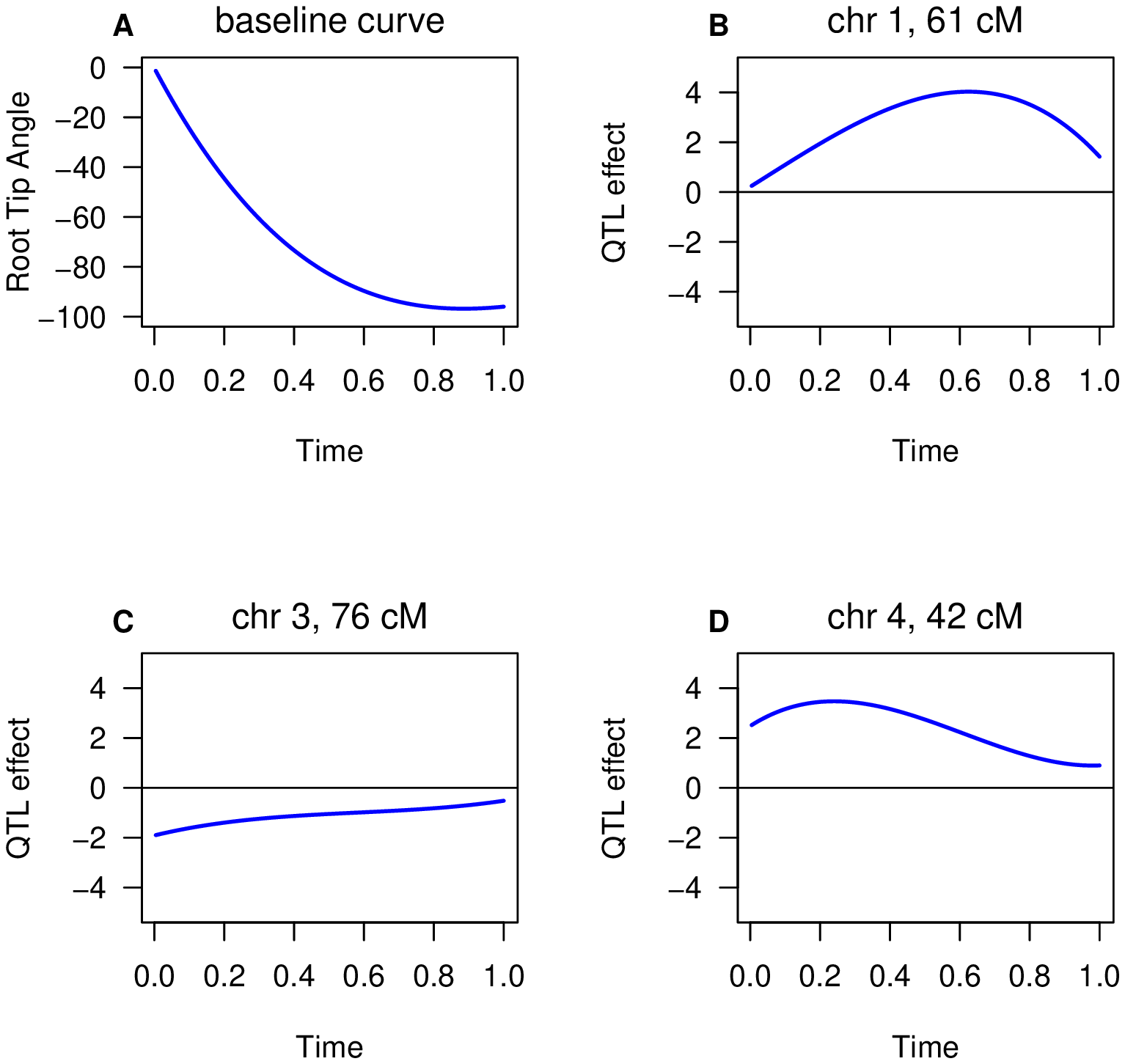}
\vspace{1cm}
 \caption{The underlying true baseline function (A) and
   QTL effect curves (B, C and D) for the multiple-QTL simulations.}
\end{center}
\end{figure}

\clearpage

\begin{table}[p]

\centering
  \caption{5\% significance thresholds for the data from Moore \emph{et al.} 2013, based on a permutation test with 1000 replicates.}

\bigskip

    \begin{tabular}{l@{\hspace{6mm}}cr@{.}lc}
      \hline
      \textbf{Method} & \multicolumn{4}{c}{\textbf{Threshold}} \\
      \hline
      SLOD         &\hspace{1mm} & 1&85   &\hspace{1mm} \\
      MLOD         && 3&32   &\\
      EE(Wald)     && 5&72   &\\
      EE(Residual) && 0&0559 &\\
      \hline
    \end{tabular}
\end{table}

\clearpage

\begin{table}[p]
\centering
  \caption{The unstructured covariance matrix used in the single-QTL simulations.}
\begin{equation*}
  \Sigma = \begin{pmatrix}
0.72 & 0.39 & 0.45 & 0.48 & 0.50 & 0.53 & 0.60 & 0.64 & 0.68 & 0.68 \\
0.39 & 1.06 & 1.61 & 1.60 & 1.50 & 1.48 & 1.55 & 1.47 & 1.35 & 1.29 \\
0.45 & 1.61 & 3.29 & 3.29 & 3.17 & 3.09 & 3.19 & 3.04 & 2.78 & 2.53 \\
0.48 & 1.60 & 3.29 & 3.98 & 4.07 & 4.01 & 4.17 & 4.18 & 4.00 & 3.69 \\
0.50 & 1.50 & 3.17 & 4.07 & 4.70 & 4.68 & 4.66 & 4.78 & 4.70 & 4.36 \\
0.53 & 1.48 & 3.09 & 4.07 & 4.68 & 5.56 & 6.23 & 6.87 & 7.11 & 6.92 \\
0.60 & 1.55 & 3.19 & 4.17 & 4.66 & 6.23 & 8.59 & 10.16 & 10.80 & 10.70 \\
0.64 & 1.47 & 3.04 & 4.18 & 4.78 & 6.87 & 10.16 & 12.74 & 13.80 & 13.80 \\
0.68 & 1.35 & 2.78 & 4.00 & 4.70 & 7.11 & 10.80 & 13.80 & 15.33 & 15.35 \\
0.68 & 1.29 & 2.53 & 3.69 & 4.36 & 6.92 & 10.70 & 13.80 & 15.35 & 15.77 \\
  \end{pmatrix}
\end{equation*}
\end{table}

\end{document}